\documentclass{jpsj3}
\usepackage{txfonts}
\usepackage{color}

\title{Development of spatial inhomogeneity of internal magnetic field above $T_{\rm c}$ in Bi$_2$Sr$_2$Ca$_{1-x}$Y$_x$Cu$_2$O$_{8+\delta}$ observed by longitudinal-field muon-spin-relaxation}

\author{Yoichi Tanabe$^1$\thanks{Present address: Department of Physics, Graduate School of Science, Tohoku University, 6-3 Aoba, Aramaki, Aoba-ku, Sendai 980-8578, Japan}, Tadashi Adachi$^1$\thanks{E-mail:t-adachi@sophia.ac.jp,{\,}Present address: Department of Engineering and Applied Sciences, Faculty of Science and Technology, Sophia University, 7-1 Kioi-cho, Chiyoda-ku, Tokyo 102-8554, Japan}, Kensuke M. Suzuki$^1$, Megumi Akoshima$^1$\thanks{Present address: National Institute of Advanced Industrial Science and Technology (AIST), 1-1-1 Umezono, Tsukuba 305-8568, Japan}, Satoshi Heguri$^2$, \\ Takayuki Kawamata$^3$\thanks{Present address: Department of Applied Physics, Graduate School of Engineering, Tohoku University, 6-6-05 Aoba, Aramaki, Aoba-ku, Sendai 980-8579, Japan}, Yasuyuki Ishii$^3$\thanks{Present address: Department of Physics, Tokyo Medical University, 6-1-1 Shinjuku, Shinjuku-ku, Tokyo 160-8402, Japan}, Takao Suzuki$^3$\thanks{Present address: College of Engineering, Shibaura Institute of Technology, 307 Fukasaku, Minuma, Saitama 337-8570, Japan}, Isao Watanabe$^3$ and Yoji Koike$^1$}

\inst{$^1$Department of Applied Physics, Graduate School of Engineering, Tohoku University, 6-6-05 Aoba, Aramaki, Aoba-ku, Sendai 980-8579, Japan \\
$^2$WPI-Advanced Institutes of Materials Research, Tohoku University, Aoba, Aramaki, Aoba-ku, Sendai 980-8577, Japan \\
$^3$Advanced Meson Science Laboratory, Nishina Center for Accelerator-Based Science, Institute of Physical and Chemical Resaerch (RIKEN), 2-1 Hirosawa, Wako 351-0198, Japan}

\abst{Longitudinal-field muon-spin-relaxation measurements have revealed inhomogeneous distribution of the internal magnetic field at temperatures above the bulk superconducting (SC) transition temperature, $T_{\rm c}$, in slightly overdoped Bi$_2$Sr$_2$Ca$_{1-x}$Y$_x$Cu$_2$O$_{8+\delta}$. The distribution width of the internal magnetic field, $\Delta$, evolves continuously with decreasing temperature toward $T_{\rm c}$. The origin of the increase in $\Delta$ is discussed in terms of the creation of SC domains in a sample.}

\kword{High-$T_{\rm c}$ cuprates, Superconducting domain, Muon spin relaxation, Bi$_2$Sr$_2$Ca$_{1-x}$Y$_x$Cu$_2$O$_{8+\delta}$, Longitudinal field}

\begin{document}
\maketitle

\section{Introduction}
One of intriguing concepts in the high-$T_{\rm c}$ superconductors is the possible formation of Cooper pairs at temperatures above the bulk superconducting (SC) transition temperature, $T_{\rm c}$.
In the high-$T_{\rm c}$ cuprates, the exchange interaction between Cu spins, $J$ $\sim$ 1000 K, which can be related to the pairing interaction, is higher than the energy scale of the pairing interaction in the conventional superconductors \cite{Uemura1, Uemura2}.
In this case, phase-fluctuating superconductivity is expected to be realized at temperatures above $T_{\rm c}$.
Both the Nernst coefficient and the torque magnetometory in La$_{2-x}$Sr$_x$CuO$_4$ (LSCO), Bi$_2$Sr$_2$CuO$_{6+\delta}$, Bi$_2$Sr$_2$CaCu$_2$O$_{8+\delta}$ (BSCCO), YBa$_2$Cu$_3$O$_{7-\delta}$ (YBCO) have suggested that the so-called vortex-liquid state takes place at temperatures far above $T_{\rm c}$ \cite{Wang1, Wang2, Wang3}.
The in-plane resistivity study of the upper critical field in the overdoped regime of LSCO has revealed that the onset temperature of the SC fluctuation is in excellent agreement with those estimated by the Nernst coefficient and the torque magnetometory \cite{Hussey}.
On the other hand, high-frequency conductivity measurements in LSCO, BSCCO, YBCO, and HgBa$_2$CuO$_{4+\delta}$ have revealed that the SC fluctuation is confined to a narrow temperature-range of 1 - 18 K just above $T_{\rm c}$, suggesting that the origin of the enhancement of the Nernst signal is different from the SC fluctuation \cite{Corson1, Maeda1, Maeda3, Maeda4, Bosovic1, Grbic1, Grbic2}.
Theoretical works have suggested that the superconductivity develops in a form of small domains at temperatures far above $T_{\rm c}$ and that the increase in the number of the small SC domains brings about the bulk SC state owing to percolation or Josephson coupling \cite{Alvarez, Kresin, Mello}.

The recent study of the scanning tunneling microscopy (STM) in BSCCO has suggested that pairing gaps tend to nucleate in nanoscale regions at temperatures far above $T_{\rm c}$ \cite{Gomes, Pasupathy, Parker}.
Moreover, both the linear diamagnetic response in the magnetization measurements in Tl$_2$Ba$_2$CuO$_{6+\delta}$\cite{Bergemann, Geshkenbein}, the hysteresis in the low-field magnetization curve in LSCO \cite{Panagopoulos}, the enhancement of the relaxation rate in the transverse-field (TF) muon-spin-relaxation ($\mu$SR) measurements in LSCO and YBCO \cite{Sonier} and the result of high-frequency conductivity measurements in LSCO \cite{Maeda1, Maeda3, Maeda4} have also suggested the presence of small SC domains at temperatures above $T_{\rm c}$.

In this paper, we have performed longitudinal-field (LF) $\mu$SR measurements in Bi$_2$Sr$_2$Ca$_{1-x}$Y$_x$Cu$_2$O$_{8+\delta}$ to detect the small SC domains at temperatures above $T_{\rm c}$.
If the small SC domains are developed locally, magnetic fluxes are expelled from the small SC domains due to the Meissner effect in low magnetic fields, giving rise to spatial inhomogeneity of the internal magnetic field.
The LF-$\mu$SR is a potential technique to detect the local inhomogeneity of the static internal field through the distribution width of the static internal field at each muon site, $\Delta$, which is similar to the TF-$\mu$SR.
We have observed in slightly overdoped Bi$_2$Sr$_2$Ca$_{1-x}$Y$_x$Cu$_2$O$_{8+\delta}$ with $x$ = 0.2 ($T_{\rm c}$ = 81 K) that $\Delta$ increases with decreasing temperature below 135 K in a low field of 20 Oe.
The relation between the increase in $\Delta$ and the emergence of small SC domains is discussed.

\section{Experimental}
Polycrystalline samples of Bi$_2$Sr$_2$Ca$_{1-x}$Y$_x$Cu$_2$O$_{8+\delta}$ with $x$ = 0 and 0.2 were prepared by the usual solid-state reaction method \cite{Akoshima}.
All samples were checked by the powder x-ray diffraction measurements to be of the single phase.
Electrical-resistivity measurements were carried out using the four-probe method.
 Values of $T_{\rm c}$, defined at the midpoint of the SC transition in the electrical resistivity, are 72 K and 81 K for $x$ = 0 and 0.2, respectively.

Zero-field (ZF) and LF-$\mu$SR measurements were performed at the RIKEN-RAL Muon Facility at the Rutherford-Appleton Laboratory in the UK using a pulsed muon beam \cite{Matsuzaki}. 
The LF was applied along the initial muon polarization.
The asymmetry parameter at a time $t$, $A$($t$), is given by $A$($t$) = ($F$($t$) - $\alpha$$B$($t$))/($F$($t$) + $\alpha$$B$($t$)), where $F$($t$) and $B$($t$) are total muon events of the forward and backward counters aligned in the beam line, respectively.
The $\alpha$ is a calibration factor reflecting relative counting efficiencies between the forward and backward counters.
The $\mu$SR time spectrum, namely, the time evolution of $A$($t$) was measured down to 40 K to monitor the evolution of small SC domains at high temperatures above $T_{\rm c}$.

\section{Results}

Figure 1 shows the $\mu$SR time spectra for Bi$_2$Sr$_2$Ca$_{1-x}$Y$_x$Cu$_2$O$_{8+\delta}$ with $x$ = 0 and 0.2.
In ZF,  the spectra show Gaussian-type slow depolarization at 200 K due to paramagnetic nuclear spins giving rise to static internal magnetic field with the Gaussian distribution at each muon site in the $\mu$SR time window.
The Gaussian-type depolarization is well fitted using the Kubo-Toyabe function \cite{KT}.
All the spectra show Gaussian-type slow depolarization more or less in ZF.
In LF of 20 Oe, the spectra tend to be decoupled and a weak oscillation is observed at 200 K, which is described by the Kubo-Toyabe function in LF.
The internal magnetic field at each muon site is given by the sum of LF and the nuclear dipole field.
In this case, muon spins precess under respective internal magnetic fields, resulting in the oscillation in the $\mu$SR time spectrum.
The oscillation in LF is marked in the bulk SC state below 40 K for $x$ = 0.2 and below 60 K for $x$ = 0 where the superconductivity is confirmed in the electrical-resistivity measurements.


In order to analyze the LF-$\mu$SR time spectra, the following equation has been used,

\begin{equation}
A(t) = A_0 e^{-\lambda t} G_Z(\Delta,H_{\rm LF},t).
\label{eq1}
\end{equation}
Here, $A_0$ and $\lambda$ are the initial asymmetry and the dynamical depolarization rate of muon spins, respectively.
The $G_Z$($\Delta$,$H_{\rm LF}$,$t$) is the static Kubo-Toyabe function in LF \cite{KT}.
The $H_{\rm LF}$ is LF.
The time spectra are well fitted with Eq. (1), as shown by solid lines in Figs. 1 (c) and (d).

Figure 2 shows the temperature dependence of $\Delta$ and $\lambda$ for $x$ = 0 and 0.2.
In the analysis of the LF-$\mu$SR time spectra, following three conditions were used;
(i) Both $\Delta$ and $\lambda$ were treated as free parameters, 
(ii) $\Delta$ was a free parameter and $\lambda$ = 0, 
(iii) $\Delta$ was a free parameter and $\lambda$ was the averaged value of $\lambda$ obtained on the condition (i) over the entire temperature range.
On the condition (i), the depolarization of muon spins is due to both dynamical and static internal magnetic fields at each muon site.
On the conditions (ii) and (iii), the dynamical depolarization of muon spins is neglected and is assumed to be independent of temperature, respectively.
For both samples, in ZF, the temperature dependence of $\lambda$ is found to be almost flat, while $\Delta$ exhibits a gradual increase with decreasing temperature regardless of the analysis condition.
In LF of 20 Oe, $\Delta$ gradually increases below $\sim$ 135 K for $x$ = 0.2 and then it rapidly increases below $T_{\rm c}$ for both $x$ = 0 and 0.2 regardless of the analysis condition.
The $\lambda$ is almost independent of temperature and values of $\lambda$ are smaller in LF than those in ZF.

\section{Discussion}
It is found in LF of 20 Oe that $\Delta$ gradually increases with decreasing temperature below $\sim$ 135 K for $x$ = 0.2 and below $\sim$ 72 K for $x$ = 0, though error bars of $\Delta$ for $x$ = 0.2 are a little bit large to determine the onset temperature of the increase in $\Delta$,  $T_{\mu}$, precisely.
Here, we discuss the origin of the increasing $\Delta$.

\subsection{Effects on $\Delta$ of the Cu-spin correlation, Curie-like paramagnetism, muon diffusion and voltex-liquid state}
First, it is noted that the present results have no relation to the development of the Cu-spin correlation.
In the underdoped regime of LSCO, it has been observed that $\lambda$ increases and $\Delta$ decreases with decreasing temperature in ZF \cite{Panagopoulos2, Adachi, Watanabe} and in LF of 20 Oe \cite{Tanabe} at low temperatures due to the slowing down of the Cu-spin fluctuation in the $\mu$SR time window.
Present samples of Bi$_2$Sr$_2$Ca$_{1-x}$Y$_x$Cu$_2$O$_{8+\delta}$ with $x$ = 0 and 0.2 exhibit no increase in $\lambda$ with decreasing temperature and reside in the slightly overdoped regime, where no slowing down of the Cu-spin fluctuation is observed in the $\mu$SR time window in ZF \cite{Panagopoulos2}.
Therefore, the present results are irrelevant to the development of the Cu-spin correlation.

Similarly, as a source of the increasing $\Delta$ above the bulk $T_{\rm c}$ for $x$ = 0.2 in LF of 20 Oe, one may think possible effects of fluctuating magnetic moments of electrons giving rise to Curie-like paramagnetism observed in the overdoped cuprates\cite{Oda, Tamegai}.
Supposed the increasing $\Delta$ above $T_{\rm c}$ for $x$ = 0.2 is due to the Curie-like paramagnetism, a similar or more pronounced behavior must be observed for $x$ = 0, because the Curie-like paramagnetism becomes marked with increasing hole-concentration in the overdoped regime\cite{Oda, Tamegai}.
Apparently, this is not the case above $T_{\rm c}$ for $x$ = 0.
Therefore, the present results in LF of 20 Oe are irrelevant to fluctuating magnetic moments of electrons.

As for the thermal diffusion of muons, the damping of $\Delta$ due to the thermal activation obeys the Arrhenius law,
\begin{equation}
\Delta \sim exp(E_{\rm A}/k_{\rm B}T).
\label{eq2}
\end{equation}
Here, $E_{\rm A}$ is the activation energy and $k_{\rm B}$ is the Boltzmann constant.
Figure 3 shows the Arrhenius plot of the temperature dependence of $\Delta$ for $x$ = 0 and 0.2.
It is found that the ln$\Delta$ increases with decreasing temperature and that the slope becomes small gradually below 140 K in ZF, suggesting that muons tend to stop the diffusion in a sample below $\sim$ 140 K.
However, this contradicts with the gradual increase in $\Delta$ in $H_{\rm LF}$ = 20 Oe for $x$ = 0.2 below 135 K as shown in Fig. 3 (c).
Therefore, this is not the case.
It is noted that the quantum diffusion of muons in the SC state is not related with the present results, because $\Delta$ smoothly develops around $T_{\rm c}$ in ZF for $x$ = 0 and 0.2 \cite{Kadono}.

Finally, it is noted that the present results are not related to the vortex-liquid state above $T_{\rm c}$, either,  \cite{Wang1, Wang2, Wang3} since vortices fluctuate with the characteristic time of 10$^{-11}$ s in high-$T_{\rm c}$ cuprates\cite{Sonier}, which is shorter than the $\mu$SR time window, i.e., 10$^{-5}$ - 10$^{-9}$ s.

\subsection{Possible emergence of superconducting domains above $T_{\rm c}$}
As shown in Fig. 2, $\Delta$ rapidly increases below $T_{\rm c}$ in $H_{\rm LF}$ = 20 Oe on field cooling for $x$ = 0 and 0.2, while no rapid increase in $\Delta$ is observed in ZF.
For $x$ = 0.2, $\Delta$ gradually increases below $\sim$ 135 K.
It has theoretically and experimentally been suggested that small SC domains are locally generated in the CuO$_2$ plane at temperatures far above $T_{\rm c}$ and that percolation or Josephson coupling between them induces the bulk SC state at $T_{\rm c}$ \cite{Maeda1, Maeda3, Alvarez, Kresin, Mello, Gomes, Pasupathy, Parker, Bergemann, Geshkenbein, Panagopoulos, Sonier}.
Supposed that a few nanoscale SC domains are developed in the CuO$_2$ plane, magnetic fluxes are expelled from the nanoscale SC domains due to the Meissner effect.
In this case, the internal magnetic field parallel to the initial muon polarization inside the SC domains decrease and the bending of the magnetic fluxes generates the internal magnetic field perpendicular to the initial muon polarization.
Moreover, when the bulk SC state is developed below $T_{\rm c}$, the expulsion of magnetic fluxes becomes strong and some vortices are randomly pinned down at the grain boundary of the polycrystalline sample on field cooling.
In this case, it is expected that the spatial inhomogeneity of the internal magnetic field parallel and perpendicular to the initial muon polarization is enhanced.
In the present analysis, since $H_{\rm LF}$ is fixed to be 20 Oe, the decrease in the internal magnetic field parallel to the initial muon polarization as well as the increase in the internal magnetic field perpendicular to the initial muon polarization could contribute to the enhancement of $\Delta$.
In high-$T_{\rm c}$ cuprates, the typical value of the penetration depth is in a micrometer scale.
Since SC domains are three orders of magnitude smaller in size than the penetration depth, the decrease in the internal magnetic field parallel to the initial muon polarization is negligibly small.
Quantitatively speaking, the transverse muon spin relaxation rate, $\sigma$, can approximate to the following equation \cite{Brandt},

\begin{equation}
\sigma \sim 3.4 \times 10^4 (\lambdaup_{\rm SC}^{-2})(1 - (H_{\rm LF}/H_{\rm c2})). \label{eq1}
\end{equation}
Here, $\lambdaup$$_{\rm SC}$ is the penetration depth and $H_{\rm c2}$ is the upper critical field.
Assuming $\lambdaup$$_{\rm SC}$ $\sim$ 1 $\mu$m and $H_{\rm LF}$ $\ll$ $H_{\rm c2}$, $\sigma$ is $\sim$ 0.034 $\mu$sec$^{-1}$ in the weak limit of magnetic field.
On the other hand, the effective $\lambdaup$$_{\rm SC}$ would increase with increasing temperature, especially, in the vicinity of the onset temperature of superconductivity.
In addition, not only the suppression of the magnetic field component perpendicular to the initial muon polarization due to the bending of magnetic fluxes around SC domains but also the motional narrowing caused by the vortex motion reduces $\sigma$.
If the present results are related to the emergence of SC domains above $T_{\rm c}$, it may be the case that the increasing number of SC domains results in the overlap between SC domains, leading to the increase in $\Delta$ in LF.
In this case, one may expect to detect the development of SC domains from the Meissner diamagnetism, because the size of overlapped SC domains could be comparable to $\lambdaup$$_{\rm SC}$.
From the $\chi$ measurement for $x$ = 0.2, however, no clear development of the Meissner diamagnetism was observed below $T_{\rm \mu}$, although the thermal hysteresis of $\chi$ observed below 180 K, probably relating to the opening of the psuedogap \cite{Pana3, Pana4}, might make the onset of the Meissner diamagnetism unclear below $T_{\rm \mu}$ \cite{chiT}.

In Fig. 4, $T_{\rm \mu}$ is plotted as a function of the hole concentration, $p$, together with $T_{\rm c}$ \cite{Akoshima}.
The onset temperature of the inhomogeneous superconductivity obtained by STM, $T_{\rm c}$$^{\rm STM}$, in Bi$_2$Sr$_2$CaCu$_2$O$_{8+\delta}$ are also plotted \cite{Gomes}.
The $p$ value has been determined using the empirical relation between $T_{\rm c}$ and $p$ in the high-$T_{\rm c}$ cuprates \cite{Tallon}.
The value of $T_{\rm \mu}$ is 135 $\pm$ 19 K far above $T_{\rm c}$ at $p$ = 0.19 ($x$ = 0.2), while it is $\sim$ 72 K, namely, around $T_{\rm c}$ at $p$ = 0.21 ($x$ = 0).
These values of $T_{\rm \mu}$ are found to roughly coincide with $T_{\rm c}$$^{\rm STM}$ values in the overdoped regime.
Although the relatively large error of $\Delta$ prevent one from obtaining the exact value of $T_{\mu}$, the development of small SC domains may be one possible explanation for the increasing $\Delta$ above $T_{\rm c}$.
Since the quantitative understanding of the possible creation of small SC domains below $T_{\rm \mu}$ has not yet been obtained, further experiments using different methods are necessary to understand the origin of the gradual increase in $\Delta$ above $T_{\rm c}$.

\section{Summary}
We have performed LF-$\mu$SR measurements in Bi$_2$Sr$_2$Ca$_{1-x}$Y$_x$Cu$_2$O$_{8+\delta}$ to detect possible small SC domains at temperatures above $T_{\rm c}$ suggested theoretically and experimentally \cite{Maeda1, Maeda3, Alvarez, Kresin, Mello, Gomes, Pasupathy, Parker, Bergemann, Panagopoulos, Geshkenbein, Sonier}.
Regardless of the analysis condition, a gradual increase in the distribution width of the static internal field at each muon site has been observed at temperatures above $T_{\rm c}$ for $x$ = 0.2.
This gradual increase in $\Delta$ at temperatures above $T_{\rm c}$ could not be understood to be due to the development of Cu-spin correlation, Curie-like paramagnetism, diffusion of muons nor voltex motions.
For the scenario of the emergence of nanoscale SC domains above $T_{\rm c}$, a rough quantitative estimation of $\Delta$ shows that the increase in $\Delta$ corresponds to the increasing number of SC domains in which the overlap between SC domains takes place.
Although $T_{\rm \mu}$ has been found to roughly coincide with $T_{\rm c}$$^{\rm STM}$ in the overdoped regime, further measurements with higher precision are necessary to conclude the possible development of SC domains above $T_{\rm c}$.

\section*{Acknowledgements}
The authors are grateful to A. Koda and M. Miyazaki for their useful comments.
The $\mu$SR measurements at the RIKEN-RAL Muon Facility were partially supported by Global COE Program gMaterials Integration International Center of Education and Research, Tohoku University,h MEXT, Japan, and by gEducation Program for Biomedical and Nano-Electronics, Tohoku University'' Program for Enhancing Systematic Education in Graduate Schools, MEXT, Japan.
One of the authors (Y. T.) was supported by the Japan Society for the Promotion of Science.

\clearpage

\begin{figure*}[htbp]
\includegraphics[width=0.8\linewidth]{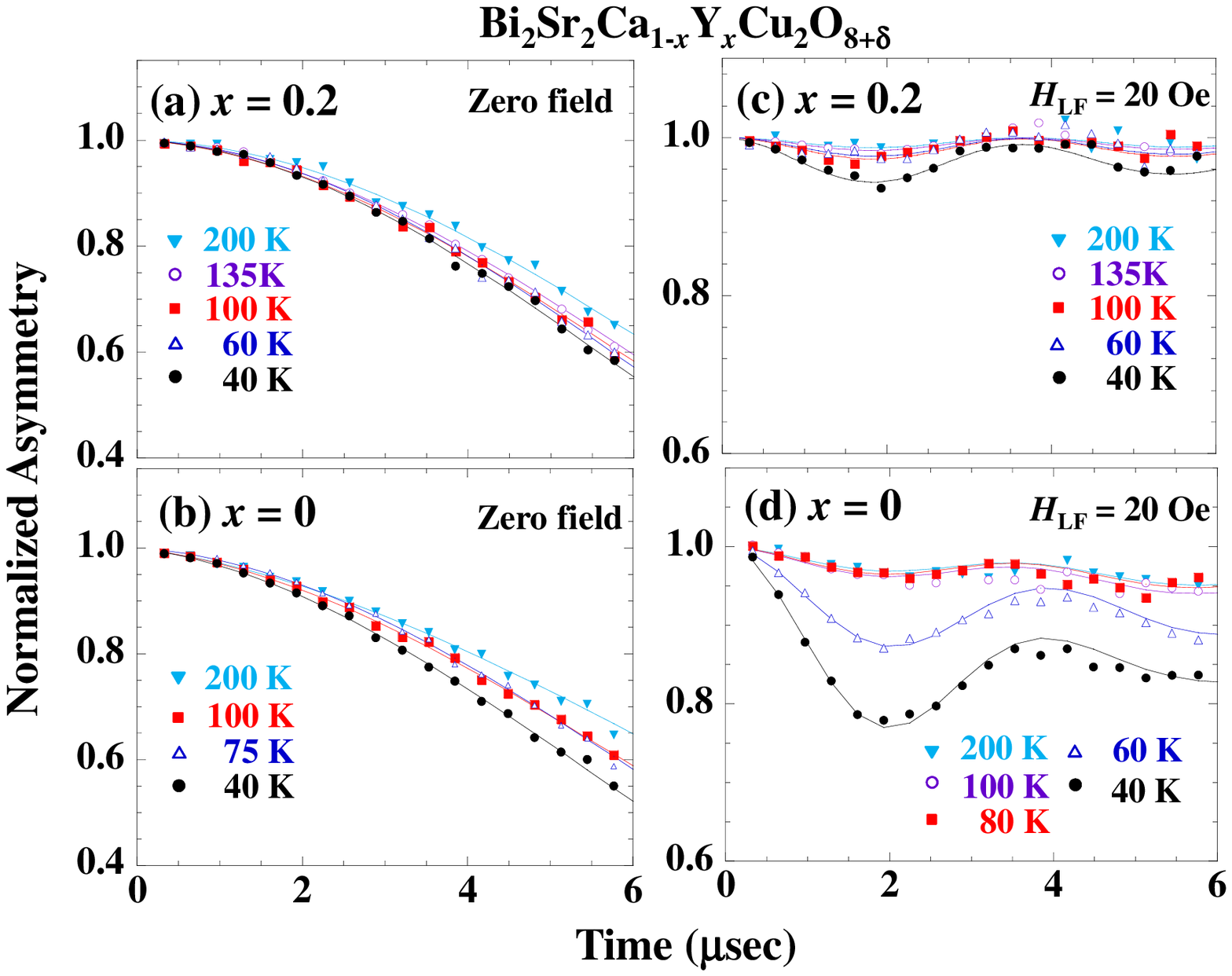}
\caption{(a), (b) Zero-field and (c), (d) longitudinal-field $\mu$SR time spectra of Bi$_2$Sr$_2$Ca$_{1-x}$Y$_x$Cu$_2$O$_{8+\delta}$ with $x$ = 0.2 and 0 at various temperatures from 200 K down to 40 K. Solid lines indicate the best-fit results using $A(t) = A_0 e^{-\lambda t} G_{\rm Z}(\Delta,H_{\rm LF},t)$.}
\end{figure*}

\clearpage

\begin{figure*}[htbp]
\begin{center}
\includegraphics[width=1.0\linewidth]{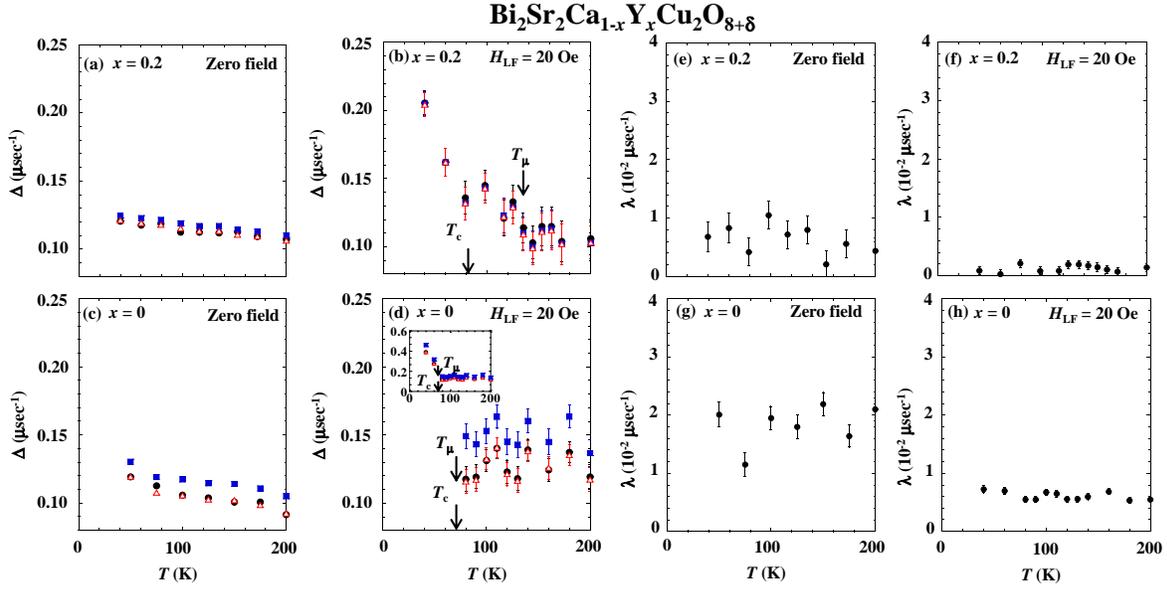}
\caption{Temperature dependences of (a) - (d) the distribution width of the static internal field at each muon site, $\Delta$, and (e) - (h) the dynamical muon-spin depolarization rate, $\lambda$, in zero field and in a longitudinal field of $H_{\rm LF}$ = 20 Oe in Bi$_2$Sr$_2$Ca$_{1-x}$Y$_x$Cu$_2$O$_{8+\delta}$ with $x$ = 0.2 and 0 obtained from the fit of the ZF- and LF-$\mu$SR time spectra with the equation $A(t) = A_0 e^{-\lambda t} G_{\rm Z}(\Delta,H_{\rm LF}$, $t$). 
The inset of (d) is the overall view of (d).
Black circles indicate the case that both $\Delta$ and $\lambda$ were free parameters.
Blue squares indicate the case that $\Delta$ was a free parameter and $\lambda$ = 0.
Red triangles indicate the case that $\Delta$ was a free parameter and $\lambda$ was the averaged value of $\lambda$, obtained on the condition that both $\Delta$ and $\lambda$ were free parameters over the entire temperature range.
The onset temperature of the increase in $\Delta$ with decreasing temperature, $T_{\rm \mu}$, and $T_{\rm c}$, defined at the midpoint of the superconducting transition in the electrical resistivity, are indicated by arrows.}
\end{center}
\end{figure*}

\clearpage

\begin{figure*}[htbp]
\includegraphics[width=0.65\linewidth]{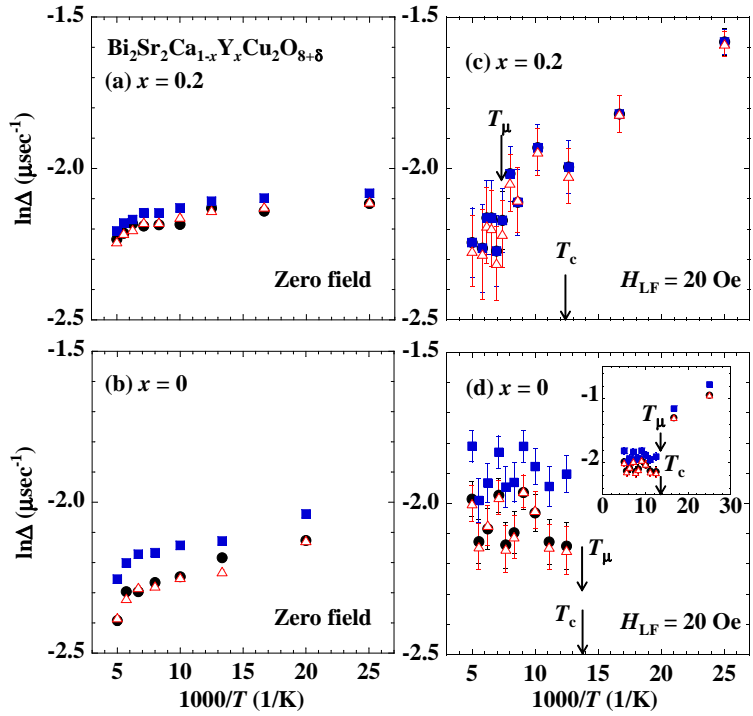}
\caption{Arrhenius plot of the temperature dependence of the distribution width of the static internal field at each muon site, $\Delta$, in zero field for (a) $x$ = 0.2 and (b) $x$ = 0 and in $H_{\rm LF}$ = 20 Oe for (c) $x$ = 0.2 and (d) $x$ = 0. Black circles, blue squares, and red triangles correspond to $\Delta$ estimated on the analysis conditions shown in the caption of Fig. 2.}
\end{figure*}

\clearpage

\begin{figure}[htbp]
\includegraphics[width=0.65\linewidth]{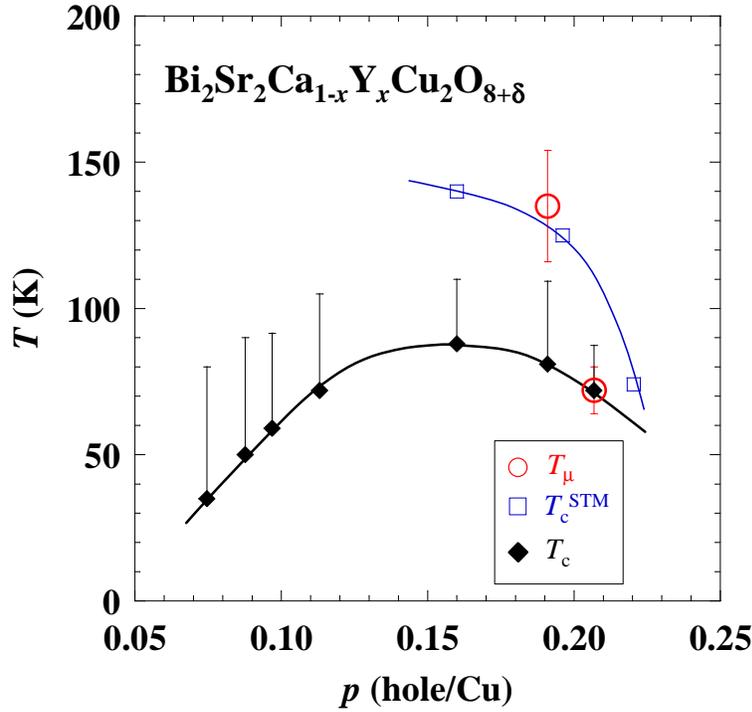}
\caption{The hole-concentration, $p$, dependence of the onset temperature of the rapid increase in $\Delta$ with decreasing temperature in LF of 20 Oe, $T_{\rm \mu}$, (red cricles), and the bulk superconducting transition temperature, $T_{\rm c}$, (black diamonds) \cite{Akoshima}. The error bar of $T_{\rm \mu}$ has been determined from the temperature dependence of $\Delta$ shown in Fig. 2(b) and (d). 
The error bar of $T_{\rm c}$ has been estimated from the onset of the superconducting transition in the electrical resistivity.
The onset temperature of the inhomogeneous superconductivity obtaine by the scanning tunneling microscope, $T_{\rm c}$$^{\rm STM}$, (open squares) in Bi$_2$Sr$_2$CaCu$_2$O$_{8+\delta}$ is also plotted \cite{Gomes}.
The $p$ value has been determined using the empirical relation between $T_{\rm c}$ and $p$ in the high-$T_{\rm c}$ cuprates \cite{Tallon}. 
Lines are guide for eyes.}
\end{figure}

\end{document}